Technische Universität Braunschweig

Carl-Friedrich-Gauß-Fakultät, Department Informatik

Institut für Software Systems Engineering

# System Model Semantics of Class Diagrams

Informatik-Bericht 2008-05

**María Victoria Cengarle**[1], **Hans Grönniger**[2]
and **Bernhard Rumpe**[2]
with the help of
**Martin Schindler**[2]

[1]Software and Systems Engineering,
Technische Universität München, Germany
[2]Software Systems Engineering,
Technische Universität Braunschweig, Germany

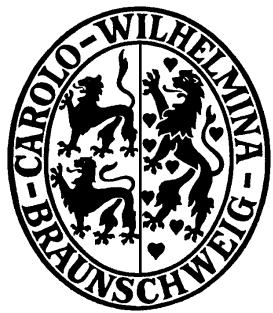

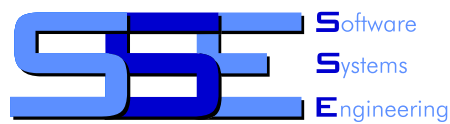

July 8, 2008

# Contents

Defining semantics for UML is a difficult task. Disagreement in the meaning of UML constructs as well as the size of UML are major obstacles. In this report, we describe our approach to define the semantics for UML. Semantics is defined denotationally as a mapping into our semantics domain called the system model [4, 5, 6]. We demonstrate our approach by defining the semantics for a comprehensive version of class diagrams. The semantics definition is detailed for UML/P class diagrams, a variant of class diagrams which restricts the use of a few methodologically and semantically involved concepts. Class diagrams are well-known and rather easy to understand and thus perfect to examine the usability of the system model for precise semantic mappings.

# 1 Introduction

Managed by the Object Management Group (OMG), UML [29, 30] offers an industry-standard means of modeling complex systems. It is a unification of several leading object-oriented modeling approaches from the mid 1990s. Modeling is the task of creating representations of various aspects of a software system prior to building it. The models convey information about the system from a variety of perspectives. UML offers a number of diagrammatic notations, intended for intuitive understanding of the model of interest. Unfortunately, the current situation is that UML still lacks a commonly agreed upon, precise semantics that allows for unambiguously validating and analyzing its models.

One approach to define a semantics for a modeling language is to explicitly and denotationally define the kind of systems the language describes, and to clearly identify which of these systems are meant by a particular model and which are not [18].

In this report, we describe our approach to define the semantics of UML in Chapter 2. In Chapter 3, we introduce the concrete syntax of class diagrams as a MontiCore grammar which is translated to its mathematical abstract syntax in Chapter 4. Context conditions for class diagrams are stated in Chapter 5 and the semantic mapping is defined in Chapter 6. Chapter 7 discusses related work and the last chapter presents an evaluation of our approach and concludes the report.

# 2 Our Approach to Define UML Semantics

There exists a variety of approaches to define the semantics for a modeling language. We decided to follow a denotational approach instead of, e.g., an operational semantics definition. Semantics for non-executable parts of a language or for underspecified or incomplete models can be easily defined using a denotational approach but is cumbersome if not impossible in an operational approach. Our goal is to define the semantics for UML which is a modeling language used for specification of the structure and behavior of software systems and is thus not fully executable and also used for specification in early project phases where complete models of the system hardly exist.

Denotational semantics consists of three parts, see, e.g., [18]. First, a precise definition of the syntactic domain is needed, i.e., a definition of the modeling language and its context conditions. Second, a suitable semantic domain has to be developed that is capable of capturing all relevant concepts that are needed to faithfully define the semantics of all syntactic constructs. Our semantic domain is the "system model" which is a general model of object oriented, possibly distributed, systems. It especially serves as an integrated semantic domain for all UML diagram types. The system model is described separately in three technical reports, for more information on the semantic domain and additional information on the general approach to semantics, see [4, 5, 6]. Third, a definition of the semantic mapping that relates each syntactic construct of a (set of) well-formed UML diagram(s) to constructs in the system model has to be given.

The semantic mapping $sem$ is generally of the form

$$sem : UML \to \wp(SystemModel)$$

that is, the semantics of a UML document is a set of systems of the system model which reflect the properties given in the UML document.

In the following, we elaborate on the different steps that constitute a semantics definition following our approach.

## 2.1 Syntax & Context conditions

For the precise specification of the syntax of the modeling language in question, we use Monti-Core, a framework for the definition and processing of languages [22]. The syntax is defined in one or more context free grammar(s) which is an established technique to obtain precise syntax definitions. In contrast, graphical metamodeling approaches often lead to complex and hard to understand language specifications in case of complex languages (for a discussion on text-based modeling, see [17]). Note that MontiCore also supports well-known metamodeling concepts (like abstract syntax associations [22]). Multiple grammars instead of one are sometimes used to specify a language making use of MontiCore's language modularization concepts like inheritance and embedding of languages. Language inheritance means that a grammar of a language inherits all rules (or productions) from a super grammar, possibly overriding them. Language embedding allows for defining a language with "holes" which are rules that are declared as external and that are later bound to rules of a different language. Context conditions for the languages are implemented and checked on concrete models with the help of framework functionality.

The advantage of using MontiCore to define the syntax of the language is that we obtain a machine readable definition that can be used for future automated processing. For instance, since MontiCore also contains a transformation engine, one possible application is the specification of the semantics as a transformation into theories of the theorem prover Isabelle [28] for verification purposes.

Syntax and semantics is defined for a conceptually improved variant of UML, UML/P [31, 32]. Nevertheless, advances in the standard documents [29, 30] are also considered.

## 2.2 Syntax Transformations

Frequently, the syntax of a language contains constructs that are expressible by (a number of) other constructs. This syntactic extensions often increase the usability of a language. They also provide an opportunity for syntactical transformations of a language in order to reduce the number of constructs. This step in the semantics definition is optional but has the advantage that the semantics of the transformed constructs can be explained purely on the syntactic domain. Additionally, the reduced set of constructs leads to more concise semantics mapping later on. As an example, state hierarchy in Statecharts can be removed [32]. The semantics mapping consequently only needs to deal with flat Statecharts.

In a modular language definition that uses language embedding, the reduction of syntactic constructs might lead to constraints on the embedded language. If, for instance, we want to get rid of exit actions in states in Statecharts by moving them on outgoing transition [32], and we embedded a language to specify statements or actions, this language at least has to offer an operator for the sequential composition of statements.

The transformations can be defined, tested and automated using the MontiCore transformation engine.

## 2.3 Mathematical Form of Syntax

The syntax, specified as a MontiCore grammar, is not directly suitable for the mathematical semantics mapping into the system model since it may contain technicalities and elements of concrete syntax that are cluttering up the definitions.

We therefore transform the MontiCore grammar into an equivalent mathematical abstract syntax definition using basic set theory. The equivalence is assured by careful reviews and a systematic translation.

- The optional name of a non-terminal is dropped, i.e.,

| | |
|---|---|
| MontiCore | `X = a:B b:C; Y=a:C;` |
| Maths | $X = B \times C, Y = C$ |

- Rules from a super grammar are copied to the grammar if the rule is not overridden.

- External rules are left unspecified in the grammar and can also be left unspecified in the mathematical version.

- Interface rules are defined as the union of the sets of their implementing rules. i.e.,

| | |
|---|---|
| MontiCore | `interface A; X implements A; Y implements A;` |
| Maths | $A = X \cup Y$ |

- Optional elements `X?` become sets that are assumed to contain a specific element that represents the non-existence, we abbreviate as follows $X^{opt} = X \cup \{\epsilon\}$ where $\epsilon$ is a special element with $\epsilon \notin X$.

- Alternatives in the grammar are translated to unions of the corresponding sets, i.e.,

| | |
|---|---|
| MontiCore | `A = X \| Y;` |
| Maths | $A = X \cup Y$ |

- Repetitions in the grammar are transformed to powersets of the corresponding sets, i.e.,

| | |
|---|---|
| MontiCore | `X = (A)*;` |
| Maths | $X = \wp(A)$ |

- Lists of the form `A (',' A)*` are translated to lists $List(A)$ which is an abbreviation for a partial function $\mathbb{N} \mapsto A$ with interval $[1, \ldots, n]$ as domain.

In the presence of language inheritance, super grammars are merged with the inheriting grammar, i.e., the parent rules are copied. Embedded language parts are also left unspecified in the mathematical abstract syntax just as they are unspecified in the concrete MontiCore syntax.

As an alternative way for syntactic language transformations, we also define transformation schemes based on the mathematical abstract syntax. This offers us a way to denote complex syntax transformations in a very concise and understandable way. As a drawback, these transformation rules are not executable and can only be quality-checked by careful review. The detailed introduction of the transformation scheme on the mathematical abstract syntax and an example application of this approach, namely transformation rules for Statecharts, can be found in [8].

## 2.4 System Model Abstractions & Semantic Mapping

Frequently, a model does not characterize the whole system but focuses on specific aspects. While the underlying system model definitions remain valid, we allow specifying views and abstractions on the definitions in order to obtain a more comprehensible and focused semantics mapping. For instance, in a semantics definition for interactions, one is primarily interested in system runs that need to fulfill the properties induced by the interaction. In that case, the semantics is defined as a mapping to a view of the system model containing system runs as sequences of reachable states. Consequently, lower-level details of the systems remain hidden from the readers of the semantics.

As stated above, the semantics mapping maps a UML model to a set of possible realizations. The semantics of a set of models is then simply the intersection of their individual semantics. Further, an empty semantics yields a notion of inconsistency: there is no possible realization that fulfills the model(s).

The semantics of languages used to state conditions or actions (e.g., OCL or UML's action language) typically does not characterize whole systems but is a relation that needs a context. The semantics is given depending on a context that contains, e.g., variable bindings, the state of attributes, etc. This kind of language is also often embedded in an outer language (e.g, Statecharts that contain OCL as a language to state preconditions or invariants) which is then able to provide the required context to calculate the semantics. In the semantics mapping for the outer language, we typically do not have access to the properties of the embedded language. Hence, we have to make the assumptions that the semantics of the embedded language is defined somewhere and that the provided context information from the outer language is sufficient to calculate the semantics. This is not a limitation of the approach but a necessary consequence of the modular definition of the semantics.

# 3 Concrete Syntax of UML/P Class Diagrams

The concrete syntax of UML/P class diagrams can be found below defined as MontiCore grammars[1]. Concrete examples, and a comparison to UML class diagrams can be found in [32] which introduces this methodologically and semantically improved variant of class diagrams in detail.

The first grammar `CD.mc` is complete, only comments and some technical details regarding parser generation have been removed. The second grammar (`Common.mc`) is the super grammar of all UML/P grammars providing frequently used concepts such as stereotypes.

```
                         MontiCore-Grammar
1  package mc.umlp.cd;
2
3  /**
4  @version 1.0
5  */
6  grammar CD extends mc.umlp.common.Common {
7
8    external Value;
9    external Body;
10
11   interface CDElement;
12
13   CDDefinition =
14     Completeness?
15     Stereotype?
16     "classdiagram" Name:IDENT
17     "{"
18     (CDElements:CDElement | (Invariants:Invariant ";"))*
19     "}";
20
21   CDClass implements CDElement =
22     Completeness?
23     Modifier
24     "class" Name:IDENT
25     TypeParameters?
26     ("extends" Superclasses:ClassOrInterfaceType
27         ("," Superclasses:ClassOrInterfaceType)*)?
28     ("implements" Interfaces:ClassOrInterfaceType
29         ("," Interfaces:ClassOrInterfaceType)*)?
30     (
```

---

[1]The MontiCore UML/P grammars, version 1.0, were developed by Martin Schindler as part of the MontiCore project (www.monticore.org).

```
31        ("{"
32         (CDConstructors:CDConstructor
33         | CDMethods:CDMethod
34         | CDAttributes:CDAttribute)*
35        "}")
36        |";"
37    );
38
39  CDInterface implements CDElement =
40    Completeness?
41    Modifier
42    "interface" Name:IDENT
43    TypeParameters?
44    ("extends" Interfaces:ClassOrInterfaceType
45         ("," Interfaces:ClassOrInterfaceType)*)?
46    (
47      ("{"
48       (CDMethods:CDMethod
49       | CDAttributes:CDAttribute)*
50       "}")
51      |";"
52    );
53
54  CDEnum implements (Completeness? Modifier "enum")=>CDElement =
55    Completeness?
56    Modifier
57    "enum" Name:IDENT
58    ("implements" Interfaces:ClassOrInterfaceType
59         ("," Interfaces:ClassOrInterfaceType)*)?
60    (
61      ("{"
62         CDEnumConstants:CDEnumConstant
63               ("," CDEnumConstants:CDEnumConstant)* ";"
64         (CDConstructors:CDConstructor
65         | CDMethods:CDMethod
66         | CDAttributes:CDAttribute)*
67       "}")
68      |";"
69    );
70
71  CDEnumConstant =
72    Name:IDENT ("("
73      CDEnumParameters:CDEnumParameter
74               ("," CDEnumParameters:CDEnumParameter)*
75    ")")?;
76
77  CDEnumParameter = Value;
78
79  CDMethod =
80    Modifier
81    TypeParameters?
82    ReturnType:ReturnType
83    Name:IDENT
84    "(" (CDParameters:CDParameter ("," CDParameters:CDParameter)*)? ")"
```

```
85         ("throws" Throws:QualifiedName ("," Throws:QualifiedName)*)?
86         (Body | ";");
87
88   CDConstructor =
89     Modifier
90     TypeParameters?
91     Name:IDENT
92     "(" (CDParameters:CDParameter ("," CDParameters:CDParameter)*)? ")"
93     ("throws" Throws:QualifiedName ("," Throws:QualifiedName)*)?
94     (Body | ";");
95
96   CDParameter = Type Name:IDENT;
97
98   CDAttribute =
99     Modifier
100    Type
101    Name:IDENT
102    ("=" Value)? ";";
103
104
105  CDAssociation implements CDElement =
106    Stereotype?
107    Type:[Association:"association"|Aggregation:"aggregation"|
108         Composition:"composition"]
109    (Derived:[DERIVED:"/"])?
110    Name:IDENT?
111    LeftStereotype:Stereotype?
112    LeftCardinality:Cardinality?
113    LeftReference:QualifiedName
114    LeftQualifier:Qualifier?
115    ("(" LeftRole:IDENT ")")?
116    Arrow:[lefttoright:"->"|righttoleft:"<-"|bidirectional:"<->"|
117          simple:"--"]
118    ("(" RightRole:IDENT ")")?
119    RightQualifier:Qualifier?
120    RightReference:QualifiedName
121    RightCardinality:Cardinality?
122    RightStereotype:Stereotype? ";";
123 }
```

─────────────────────── MontiCore-Grammar ───────────────────────

```
1 package mc.umlp.common;
2
3 /**
4 @version 1.0
5 */
6 grammar Common {
7
8   ident NUMBER = ('0'..'9')+;
9
10  ident IDENT =
11    ('a'..'z' | 'A'..'Z' | '_' | '$')
```

```
12       ('a'..'z' | 'A'..'Z' | '_' | '0'..'9' | '$')*;

13
14   Stereotype =
15     "<<" Values:StereoValue ("," Values:StereoValue)* ">"">";
16
17   StereoValue = Name:IDENT ("=" Value:STRING)?;
18
19   QualifiedName = Names:IDENT "." Names:IDENT)*;
20
21   Cardinality =
22     "["
23     ( Many:["*"]
24     | LowerBound:NUMBER
25     | (LowerBound:NUMBER ".." (UpperBound:NUMBER | NoUpperLimit:["*"]))
26     )
27         "]";
28
29   Qualifier = "[" Type:ClassOrInterfaceType "]";
30
31   Modifier =
32     Stereotype?
33     ( Public:["public"] | Public:[PUBLIC:"+"]
34     | Private:["private"] | Private:[PRIVATE:"-"]
35     | Protected:["protected"] | Protected:[PROTECTED:"#"]
36     | Final:["final"]
37     | Abstract:["abstract"]
38     | Local:["local"]
39     | Derived:["derived"] | Derived:[DERIVED:"/"]
40     | Readonly:["readonly"]       | Readonly:[READONLY:"?"]
41     | Static:["static"]
42     )*;
43
44   TypeParameters =
45     "<" TypeParameters:TypeParameter
46           ("," TypeParameters:TypeParameter)* ">";
47
48   TypeParameter =
49     Name:IDENT
50     ("extends" SupTypes:ClassOrInterfaceType
51           ("&" SupTypes:ClassOrInterfaceType)*)?;
52
53   interface ReturnType;
54   interface Type;
55
56   VoidType implements ReturnType = "void";
57
58   PrimitiveType implements ReturnType, Type =
59     Primitive: [ "boolean" | "byte" | "char" | "short"
60                | "int" | "float" | "long" | "double"] ("[""]")*;
61
62   ReferenceType implements ReturnType, Type =
63     ClassOrInterfaceType ("[""]")*;
64
65   ClassOrInterfaceType =
```

```
66      Name:QualifiedName TypeArguments?;
67
68   TypeArguments =
69     "<" TypeArguments:TypeArgument
70           ("," TypeArguments:TypeArgument)* ">";
71
72   TypeArgument =
73       Type
74     | ("?" (("extends" UpperBound:ReferenceType) |
75             ("super" LowerBound:ReferenceType))?);
76
77   Invariant =
78     (Kind:IDENT ":")?
79     "[" InvariantExpression(parameter Kind) "]";
80
81   external InvariantExpression;
82
83   Completeness = ...
84
85 }
```

# 4 Abstract Syntax of UML/P Class Diagrams

We proceed with defining the mathematical abstract syntax of UML/P class diagrams. Please note, that we have refactored the syntax in the transition from the MontiCore grammars to the mathematical form. For example, we renamed rules for a more comprehensible reference later on, and we also changed details of the language as follows:

- As explained in [32, Sect. 3.4], completeness information is a syntactical means to indicate that model or model element is a view (other models might exist to complete the model) or regarded as the complete model of the system. This does not constrain the actual system from have additional behavior or structure. Hence, completeness information is irrelevant for the semantics and removed from the syntax, e.g., lines 11 and 19 in grammar `CD.mc`.

- In principle, arbitrary stereotypes can be defined for class diagrams and class diagram elements. For the semantics, we only consider the stereotypes explained in [32] and fix them in the abstract syntax. Additional stereotypes (also for other model elements) can of course later be added and their meanings incorporated in the semantics.

- Simple names are used for referring to classes, interfaces, etc. (e.g., in the list of super classes and interfaces or in associations). We assume that qualified names can be encoded somehow in these names if need be. This also means that class names have to be unique.

- Support for generic types is not considered in this version of the semantics but is a matter of future investigations.

- The interface CDElements in line 8 has been expanded as sets of classes, interfaces and associations in the mathematical syntax. This version of the class diagrams currently does not support enumeration types defined in the grammar (lines 51-74).

- We removed the modifier local as its intended use refers only to syntax: the local element may not be referenced in other diagrams. Furthermore, readonly is renamed to frozen, and we additionally consider addonly and ordered as modifiers for associations.

- Invariants in `Common.mc` (line 26) are parameterized with the name of the invariant language. This is solution to a technical problem and not needed in the mathematical version.

- Identifiers are specified in `Common.mc`, we assume a set of Names which is left unspecified.

- We currently do not support initial values for attributes.
- Cardinality in associations is fixed to the values proposed in [32].
- Types and return types as detailed in `Common.mc` are assumed to be given by some basic types (imported) and the defined class and interface types in the diagram.

Some minor changes and rearrangements such as renaming (CDDefinition to CD, IDENT to Name) will not be further detailed.

The abstract syntax for class diagrams is now given below in a mathematical form. We use the set *Name* that contains identifiers which are not further specified. The sets *Cond*, *BasicType*, and *Stmt* are also not described here but specified elsewhere. We import those parts of the language in a compositional form as described in [19]. They are part of an action language that may for example be OCL to state conditions or Java to formulate statements. We also assume some basic types, e.g., int, char etc., given and imported through *BasicType*.

$$
\begin{aligned}
\textit{CD} &= \textit{DiagramName} \times \wp(\textit{Class}) \times \wp(\textit{Interface}) \times \wp(\textit{Assoc}) \times \wp(\textit{Inv}) \\
\textit{Class} &= \wp(\textit{Modifier}) \times \textit{ClassName} \times \wp(\textit{SuperClassName}) \times \\
&\quad \wp(\textit{InterfaceName}) \times \wp(\textit{Constructor}) \times \wp(\textit{Meth}) \times \wp(\textit{Attr}) \\
\textit{Interface} &= \textit{InterfaceName} \times \wp(\textit{SuperInterfaceName}) \times \\
&\quad \wp(\textit{Meth}) \times \wp(\textit{Attr}) \\
\textit{Assoc} &= \wp(\textit{Modifier}) \times \textit{AssocName}^{opt} \times \textit{LeftPart} \times \textit{Direction} \times \textit{RightPart} \\
\textit{LeftPart}, \textit{RightPart} &= \wp(\textit{Modifier}) \times \textit{ClassName} \times \textit{Role}^{opt} \times \textit{Card}^{opt} \times \textit{Qualifier}^{opt} \\
\textit{Card} &= \{0..1, 1, *\} \\
\textit{Direction} &= \{\leftarrow, \rightarrow, \leftrightarrow, -\} \\
\textit{Constructor} &= \wp(\textit{Modifier}) \times \textit{Name} \times \\
&\quad \mathsf{List}(\textit{FormalParameter}) \times \wp(\textit{ExceptionName}) \times \textit{Body}^{opt} \\
\textit{Meth} &= \wp(\textit{Modifier}) \times \textit{Name} \times \textit{Type} \times \\
&\quad \mathsf{List}(\textit{FormalParameter}) \times \wp(\textit{ExceptionName}) \times \textit{Body}^{opt} \\
\textit{FormalParameter} &= \textit{Name} \times \textit{Type} \\
\textit{Type} &= \textit{ClassName} \cup \textit{BasicType} \cup \textit{InterfaceName} \\
\textit{Attr} &= \wp(\textit{Modifier}) \times \textit{AttrName} \times \textit{Type} \\
\textit{Modifier} &= \{\mathsf{public, private, protected, static, abstract, final, composition,} \\
&\qquad \mathsf{derived, ordered, frozen, addonly}\} \\
\textit{Qualifier} &= \textit{Type} \cup \textit{AttrName} \\
\textit{SuperInterfaceName}, & \\
\textit{SuperClassName}, & \\
\textit{ExceptionName}, & \\
\textit{ClassName}, & \\
\textit{DiagramName}, & \\
\textit{Role}, \textit{InterfaceName}, & \\
\textit{AssocName} & \\
\textit{AttrName} &= \textit{Name} \\
\textit{Inv} &= \textit{Cond} \\
\textit{Body} &= \textit{Stmt}
\end{aligned}
$$

For notational convenience, we refer to specific components of these tuples using their (dot separated) names. If the component is a set, we use the plural form of the name. E. g., given a class diagram $cd \in CD$, *cd.diagramName* is a shorthand for the projection on the first component: $cd.diagramName = \pi_1(cd)$, and *cd.classes* denotes a projection on the second: $cd.classes = \pi_2(cd)$.

# 5 Context Conditions for UML/P Class Diagrams

Not every diagram that is represented ny the above defined abstract syntax is meaningful. To be able to define a meaning, a diagram must be well-defined. The well-definedness of a class diagram $cd \in CD$ is defined through the following context conditions. We assume that the class diagram is complete in the sense that all context conditions can be checked by only using information in the diagram, so there is no need for, e.g., resolving imports from other diagrams. The list is not complete. The MontiCore implementation contains a complete coverage of context conditions for class diagrams.

1. Certain modifiers are only applicable to certain elements, namely:
    a) for classes, abstract and final,
    b) for associations, composition and derived,
    c) for (left or right) association ends, addonly, frozen and ordered,
    d) for constructors, public, private and protected,
    e) for methods, public, private, protected, abstract and static,
    f) for attributes, public, private, protected, static, final and derived.

2. Class and interface names are unique.
    $$e_1, e_2 \in (cd.classes \cup cd.interfaces) \land e_1 \neq e_2 \implies e_1.name \neq e_2.name$$

3. In the context of a class $c \in cd$, the following conditions apply.
    a) Superclasses exist in the class diagram and are not final.
       $n \in c.superClassNames \implies \exists d \in cd.classes : d.name = n \land \text{final} \notin d.modifiers$
    b) Interfaces exist in the class diagram.
       $n \in c.interfaceNames \implies \exists i \in cd.interfaces : i.name = n$
    c) The constructor name and the class name coincide.
       $k \in c.constructors \implies k.name = c.name$
    d) Attributes have all different names.
       $at_1, at_2 \in c.attrs \land at_1 \neq at_2 \implies at_1.name \neq at_2.name$
    e) There is at most one visibility modifier for attributes.
    f) Methods $m \in c.meths$ declared in a class have to respect the following conditions:
        i. There is at most one visibility modifier for methods.

ii. Exceptions are declared classes.
$n \in m.exceptionNames \implies \exists e \in cd.classes : e.name = n$

iii. Return types are declared as class or interface or are basic types.

iv. Abstract methods are only allowed in abstract classes, and have no body.
abstract $\in m.modifiers \implies$ abstract $\in c.modifiers \wedge m.body = \epsilon$

v. Formal parameters are named differently, and their types are either of basic types or else declared in the diagram if of class or interface types.

vi. Method signatures are unambiguous: for every method call, at most one method signature matches the list of actual paramenters.

vii. Visibility restrictions have to be observed. This can only be checked statically (e.g., at compile time) and depends on the choice of implementation language for methods.

g) Subclassing related conditions:

i. If the class is not abstract, the methods of an interface are implemented in the class.

ii. Method overriding rules, for instance:
$m_1 \in c_1.meths \wedge m_2 \in c_2.meths \wedge m_1.name = m_2.name \wedge$
$c_1.name \in c_2.superClassNames \wedge$
$types(m_1.formalParams) = types(m_2.formalParams) \wedge$
public $\in m_1.modifier$
$\implies$ private $\notin m_2.modifier$

4. The superclass or superinterface relationship is not circular.
$R_1 = \{(c,d) \mid c \in cd.classes \wedge d.name \in c.superClassNames\}$ and
$(a,b) \in R_1^+ \implies a \neq b$.
$R_2 = \{(c,d) \mid c \in cd.interfaces \wedge d.name \in c.superInterfaceNames\}$ and
$(a,b) \in R_2^+ \implies a \neq b$.

5. For associations $a \in cd.assocs$, the following context conditions apply.

a) An association connects declared classes.

b) Qualifiers that are attribute names correspond to attributes on the opposite class. For the left side,
$a.leftPart.qualifier = atname \wedge a.rightPart.className = cname \implies$
$\exists c \in cd : \exists at \in c.attrs : c.className = cname \wedge at.attrName = atname$

c) Analogously for the right side,
$a.rightPart.qualifier = atname \wedge a.leftPart.className = cname \implies$
$\exists c \in cd : \exists at \in c.attrs : c.className = cname \wedge at.attrName = atname$

6. If the association is of type composition then the composite (to the left) may not exceed cardinality 1.
composite $\in a.modifiers \implies a.leftCard \subseteq \{0..1, 1\}$

7. For interfaces $i \in cd.\textit{interfaces}$, we have the following conditions.

   a) Superinterfaces are declared in the class diagram.

   b) Attributes have all different names.
      $at_1, at_2 \in i.\textit{attrs} \land at_1 \neq at_2 \implies at_1.\textit{name} \neq at_2.\textit{name}$

   c) Methods of interfaces have no body.
      $m \in i.\textit{meths} \implies m.\textit{body} = \epsilon$

   d) Method signatures are unambiguous (see condition 3(f)vi above).

# 6 Mapping of UML/P Class Diagrams

A system model is said to be an implementation of a class diagram if a number of conditions hold. These conditions are divided into static and dynamic conditions. Static conditions, e.g., ensure that classes and associations declared in a class diagram can also be found in a system model. A static condition can be checked on any snapshot of a system and often much easier on the typing and structural aspects of each system. Dynamic conditions however describe behavioral aspects and therefore need to lock at system runs and object behavior. They are also needed to map syntactic concepts for which no direct semantic equivalent can be found in the system model (e.g., visibility). These are conditions that have to hold during runtime of a system model (e.g., private methods may never be called from outside the object. This is a dynamic condition that ensures that the system model respects private visibility).

Note that structural constructs that have an equivalent in the system model do normally not need dynamic conditions. The system model observes these conditions directly but could be adapted by instantiating semantic variation points. For example, given a method in the system model, the system model definitions guarantee correctness of calls regarding argument types etc. If, e.g., another notion of type safety is required for method calls, this can be formalized as a variation point with additional conditions on top of the system model definitions and need not be defined in the mapping.

**Preliminaries**

1. We assume a function $trans_t$ that translates class diagram basic types and class / interface types to system model constructs:
   $x \in \textit{BasicType} \implies trans_t(x) \in \mathsf{UTYPE}$
   $x \in (\textit{Class} \cup \textit{Interface})$
   $\implies trans_t(x) \in \mathsf{UCLASS} \wedge$
   $\quad \forall a \in x.\textit{attrs} : \exists (n, T) \in \textit{attrs}(trans_t(x)) : a.\textit{name} = n \wedge trans_t(a.\textit{type}) = T$

2. We assume a function $trans_a$ that translates class diagram associations to system model associations:
   $a \in \textit{Assoc} \implies trans_a(a) \in \mathsf{UASSOC}$
   $c_l.\textit{className} = a.\textit{leftPart.className} \wedge c_r.\textit{className} = a.\textit{rightPart.className}$
   $\implies (trans_t(c_l), trans_t(c_r)) = \textit{classesOf}(trans_a(a))$

3. Finally, we assume a function $trans_m$ that translates class diagram methods to system model methods:
   $m \in \textit{Meth} \implies trans_m(m) \in \mathsf{UMETH}$
   $trans_t(\textit{type}(m.\textit{formalParams})) = \textit{type}(\textit{parOf}(trans_m(m)))$

4. For a given system model $sys$, the reachable states are denoted as *reachableStates*($sys$).

5. With the import of a language to state conditions comes a relation as follows: $\mathcal{C}(\mathit{cond}, s)$ holds if the condition $\mathit{cond} \in \mathit{Cond}$ can be evaluated to true under the system model state $s \in \mathsf{USTATE}$.

**Mapping**

Let $cd \in CD$, let $sys$ be a system model. $sys$ is said to be a valid interpretation of $cd$ if certain static and dynamic conditions hold. The static conditions are as follows.

1. $c \in cd.classes \implies$

   a) Each class exists in the system model.
   $trans_t(c) \in \mathsf{UCLASS}_{sys}$

   b) Superclasses and interfaces in the diagram are superclasses in the system model.
   $\forall d \in cd.classes \ : \ d.name \in c.superClassNames \lor d.name \in c.interfaceNames$
   $\implies trans_t(d) \in \mathsf{UCLASS}_{sys} \land (trans(c), trans(d)) \in \mathsf{sub}_{sys}$

   c) For final classes in the diagram there are no subclasses in the system model.
   $\mathsf{final} \in c.modifiers \implies \nexists d \in \mathsf{UCLASS}_{sys} : (d, trans(c)) \in \mathsf{sub}_{sys}$

   d) $m \in c.meths \implies$

      i. For each method in the diagram there is a method in the system model.
      $trans_m(m) \in \mathsf{UMETH}_{sys}$

      ii. The method belongs to the correct class.
      $\mathsf{classOf}(trans_m(m)) = trans_t(c)$

   e) Object creation is simulated by *tsts*($\mathsf{UOID}_{sys}$).$\Delta$ if first a new *oid* is added to the data store, then the constructor is executed (or simulated) as a method of the new *oid*, and finally this *oid* is returned to the object that called the constructor.

2. $i \in cd.interfaces \implies$

   a) Each interfaces exists as a class in the system model.
   $trans_t(i) \in \mathsf{UCLASS}_{sys}$

   b) Interface methods in the diagram correspond to operations in the system model that belong to the class $trans_t(i)$ which is the translation of the interface $i$, but for which no method implementation exists that belongs to that class.
   $\forall m \in i.meths : trans_m(m) \in \mathsf{UMETH}_{sys}$
   $\implies (\mathsf{classOf}(trans_m(m)), trans_t(i)) \in \mathsf{sub}^+_{sys}$
   $\land \mathsf{classOf}(trans_m(m)) \neq trans_t(i)$

   c) Superinterfaces in the diagram are superclasses in the system model.
   $\forall d \in cd.interfaces \ : \ d.name \in i.superInterfaceNames \implies$
   $trans_t(d) \in \mathsf{UCLASS}_{sys} \land (trans(i), trans(d)) \in \mathsf{sub}_{sys}$

3. $a \in cd.assocs \implies$

   Each association in the diagram is an association in the system model.
   $trans_a(a) \in \mathsf{UASSOC}_{sys}$

The dynamic conditions are as follows.

4. $c \in cd.classes \implies$

   a) There are no instances of abstract classes.
      abstract $\in c.modifiers \implies$
      $\forall (ds, cs, es) \in reachableStates(sys) : \nexists oid \in \text{oids}(ds) : \text{classOf}(oid) = trans_t(c)$

   b) $a \in c.attrs \implies$

      i. Values of final attributes do not change.
         final $\in a.modifiers \implies$
         $\forall (ds, cs, es), (ds', cs', es') \in reachableStates(sys) :$
         $\forall oid \in oids(ds) \cap oids(ds') :$
         $\text{classOf}(oid) = trans_t(c) \implies ds(oid.a.name) = ds'(oid.a.name)$

      ii. Static attributes have the same value for every instance of the class.
          static $\in a.modifiers \implies \forall (ds, cs, es) \in reachableStates(sys) :$
          $\forall oid, oid' \in oids(ds) : \text{classOf}(oid) = \text{classOf}(oid') = trans_t(c)$
          $\implies ds(oid.a.name) = ds(oid'.a.name)$

   c) $m \in c.meths \implies$

      i. Private methods are only called by objects of the same class.
         private $\in m.modifiers \implies$
         $\forall (ds, cs, es) \in reachableStates(sys) : \forall oid \in ds(oid), \forall msg \in \text{UMESSAGE} :$
         $(\text{ReceiveEvent}(msg) \in es(oid) \wedge$
         $m.name = \text{opnOf}(msg) \wedge$
         $\text{classOf}(oid) = trans_t(c))$
         $\implies \text{classOf}(\text{sender}(msg)) = trans_t(c)$

      ii. Protected methods are only called by objects of the same class or of subclasses of that class.
          protected $\in m.modifiers \implies$
          $\forall (ds, cs, es) \in reachableStates(sys) : \forall oid \in ds(oid), \forall msg \in \text{UMESSAGE} :$
          $(\text{ReceiveEvent}(msg) \in es(oid) \wedge$
          $m.name = \text{opnOf}(msg) \wedge$
          $\text{classOf}(oid) = trans_t(c))$
          $\implies (\text{classOf}(\text{sender}(msg)), trans_t(c)) \in \text{sub}^+_{sys}$

      iii. If a system run of $sys$ simulates a method body (including exceptions), then $sys$ is a model of the mehod. The notion of method simulation is similar to that for statecharts (see [8]).

      iv. Static methods can be simulated by adding a references from every object in the system model to an object of the class defining the static method.[1]

5. $i \in cd.interfaces \implies$

---

[1] This matter is subject of future work.

There are no instances of interfaces.
$\forall (ds, cs, es) \in reachableStates(sys):$
$\nexists oid \in \mathsf{oids}(ds) : \mathsf{classOf}(oid) = trans_t(i)$

6. $c \in cd.invs \implies$

    Class diagram invariants hold.
    $\forall ((s, m), cs, es) \in reachableStates(sys) : \mathcal{C}(c, (inactive((s, m), cs, es)))$
    where $inactive((s, m), cs, es)$ is the state $((s, m), cs, es)$ exluding the active objects. More precisely, the set of inactive objects is defined by
    $s_i = \{oid \in \mathsf{UOID}_{sys} \mid cs(oid)(t) = empty \; \forall t \in \mathsf{UTHREAD}_{sys}\}$
    and thus $inactive((s, m), cs, es) = ((s_i, m), cs, es)$.

7. $a \in cd.assocs \land a.leftPart.card = \mathsf{m1..m2} \land a.rightPart.card = \mathsf{n1..n2} \implies$

    a) Multiplicity constraints are observed.
    $a.leftPart.qualifier = \epsilon \implies$
    $\forall (ds, cs, es) \in reachableStates(sys):$
    $(\forall l : \mathsf{classOf}(l) = trans_t(c) \land c.className = a.leftPart.className$
    $\implies \mathsf{n1} \leq \#\{(x, y) \in \mathsf{relOf}(trans_a(a))(ds) \mid x = l\} \leq \mathsf{n2})$
    $\land$
    $(\forall r : \mathsf{classOf}(r) = trans_t(d) \land d.className = a.rightPart.className$
    $\implies \mathsf{m1} \leq \#\{(x, y) \in \mathsf{relOf}(trans_a(a))(ds) \mid y = r\} \leq \mathsf{m2})$

    b) Qualifiers given by a type also observe the multiplicity constraints.
    $t = trans_t(a.leftPart.qualifier) \in \mathsf{UTYPE}_{sys} \implies$
    $\forall (ds, cs, es) \in reachableStates(sys) : \forall q \in \mathsf{CAR}(t):$
    $\mathsf{n1} \leq \#\{(x, y, v) \in \mathsf{relOf}(trans_a(a))(ds) \mid v = q\} \leq \mathsf{n2}$
    (And similarly for a qualifier on the right end of the association.)

    c) The same for qualifiers given by an attribute name.
    $a.leftPart.qualifier = atname \implies$
    $\forall (ds, cs, es) \in reachableStates(sys) : \forall r \in \mathsf{UOID}_{sys}:$
    $\mathsf{classOf}(r) = trans_t(c) \land c.className = a.rightPart.className \land$
    $R = \{(x, y) \in \mathsf{relOf}(trans_a(a))(ds) \mid val(ds, y, atname) = val(ds, r, atname)\}$
    $\implies \mathsf{n1} \leq \#R \leq \mathsf{n2}$
    (And similarly for a qualifier on the right end of the association.)

    d) addonly association ends can only be increased.
    $\mathsf{addonly} \in a.leftPart.modifier \lor \mathsf{addonly} \in a.rightPart.modifier \implies$
    $\forall (ds, cs, es) \in reachableStates(sys):$
    $((ds', cs', es'), out) \in \Delta((ds, cs, es), inp) \land (x, y) \in \mathsf{relOf}(trans_a(a))(ds)$
    $\implies (x, y) \in \mathsf{relOf}(trans_a(a))(ds')$

    e) frozen association ends cannot be modified.
    $\mathsf{frozen} \in a.leftPart.modifier \implies$
    $\forall (ds, cs, es) \in reachableStates(sys):$
    $((ds', cs', es'), out) \in \Delta((ds, cs, es), inp)$
    $\implies \pi_1(\mathsf{relOf}(trans_a(a))(ds)) = \pi_1(\mathsf{relOf}(trans_a(a))(ds'))$

(where $\pi_1$ is the projection on the first component).
(And similarly for a frozen right end of the association.)

f) Composition implies that the existence of the parts depends on the existence of the whole.
composition $\in a.rightPart.modifier \implies$
$\forall (ds, cs, es) \in reachableStates(sys) : \forall r \in oids(ds) :$
classOf$(r) = trans_t(c) \land c.className = a.rightPart.className \implies$
$\exists l \in oids(ds) : $ classOf$(r) = trans_t(c') \land c'.className = a.leftPart.className$
$\land (l, r) \in $ relOf$(trans_a(a))(ds)$
(And similarly for a composition modifier on the right end of the association.)

g) In case an association end is ordered, the above conditions on the realisation of the association have to be reformulated accordingly: relOf$(trans_a(a))(ds)$ contains pairs of one $oid$ and a list of $oid$s (instead of pairs of $oid$s).

A more detailed discussion of system model variation points that are primarily relevant for class diagrams (such as type system, type safety, sublcassing, associations) is subject of future work.

# 7 Related Work

There is a series of works that explore the semantics of class diagrams. The approach most closely related to the above one is [37, 38]. This work presents a denotational semantics of UML class diagrams based on transformation systems. Transformation systems extend labeled transition systems, and have much in common with our system model. To the best of our knowledge, this approach unfortunately was discontinued.

In [12, 2, 27], the expressiveness and the connected semantic complicatedness of associations are handled. The semantics of conceptual class diagrams is treated in [36], while [3] defines a semantics for refinement of associations. [33, 7, 25, 15] study the satisfiability of class diagrams with respect to the cardinality of association ends; therein, the semantics of a class diagram is understood as a set of inequations. Class diagrams are given semantics in [14] with focus on their efficient usability for configuration management.

Other approaches treat class diagrams in combination with other UML sublanguages. Tools for validation of (generated) object diagrams with respect to class diagrams with OCL constraints are presented in [16, 9]. Similarly, [26] translates class diagrams and OCL constraints into Alloy [21] with the purpose of constraint checking. In [11, 10] metamodels (UML profiles) are equipped with an operational semantics that supports semantic variation points via template parameters. For class diagrams together with statecharts, [34] particularly concentrates on refinement of associations, [13] defines an operational semantics that is compositional and also considers activity groups, [20] analyzes time as well plus verification techniques based on the Prototype Verification System (PVS). An integrated semantics for class, object and state diagrams is given in [24]; the proposal defines system evolution and is based on a graph transformations, which can be described as rewriting systems for graphs, i.e., works on the syntax of the different diagrams. A simple semantics for class and sequence diagrams based on rules reflecting properties of object-oriented programs is presented in [39], in such a way that properties proved for diagrams guarantee the executability of the model.

In general, other proposals do not support semantic variation points and/or compositionality. These two properties are a sine qua non for our purposes. The semantics for the UML has to permit different instantiations according to the particular needs of the domain of application, on the one hand. On the other, compositionality ensures that the semantics of a whole UML model, consisting of a landscape which includes views on the system-to-be expressed in different UML sublanguages, can be composed when e.g. extended, and combined when put together.

# 8 Evaluation and Conclusion

Class diagrams describe the structure or (parts of) the architecture of a system; almost any other description of the system can be based on the resulting models. That is, class diagrams are on the one hand the basic notation, and on the other need to express a wide range of aspects.

The UML 2 standard is designed to satisfy many requirements that stem from different stakeholders and application domains, and therefore necessarily is overloaded. Many of its constituent elements do not seem, in the first instance, absolutely necessary; at least not in their present form. For our purposes, thus, we resorted to UML/P [32]. UML/P is the result of a number of projects on the foundations and the applications of software engineering; details can be learned from [31, 32].

As detailed in Chapter 2, the three actors in a definition of semantics are a syntax, a semantic domain, and the relationship between them. For class diagrams one of these three actors, namely the syntax, is the one of UML/P for classes. Its formalization is done in MontiCore (see [23, 35]) and presented in Chapter 3. The semantic domain is the theory of system models, described in [4, 5, 6]. The third actor is the aim of the present work, and was presented in Chapter 6.

The structured nature of MontiCore permits the use of multiple grammars instead of a single one. Not only reuse is supported, also overriding is allowed, and moreover incomplete definitions in which the "holes" can be filled at configuration time, allowing the embedding of different languages. MontiCore grammars, therefore, can be instantiated according to different needs. In this way, semantic variation points that have an impact on the syntax can be easily dealt with. MontiCore handles context-free grammars as well as additional context conditions by means of the framework functionality.

A further advantage of MontiCore is that its grammars can be used to feed automatic tools. We plan an implementation of a UML model verification tool in Isabelle [28]. This is a major endeavor that also demands, among other things, to code the theory of system models in Isabelle.

With the objective of mapping class diagrams to system models, however, the use of a machine readable grammar is somewhat cumbersome. This is because the semantic mapping we define is not an algorithm but a number of conditions a system model has to observe in order to be declared a valid implementation of the given class diagram. Because of this reason, an equivalent form of grammar is defined in Chapter 4 based on set theory that is mathematically precise and better suited for our goal. These definitions are likewise accompanied by a number of context conditions as presented in Chapter 5.

The core contribution, detailed in Chapter 6, left some characteristics of UML/P unhandled. There are a couple of details in our definition that need further treatment, and were indicated at the corresponding places in the sections above. For some of those, like navigability, we sustain that they are matters to be checked statically (i.e., at compile time) and thus not a topic to be treated here. For some others, for instance generic types and static methods, we have an idea about their realization in the system model; they being untreated, however, does not diminish the value of this contribution. Finally, the question of a system model implementation that do not mirror one to one the names of classes and associations present in the class diagram, outstrips the purposes of the present work.

Indeed, the mapping from class diagrams to system models resulted in an effort beyond our initial expectations. This was mainly due to the fact that class diagrams that seem to constitute only a relatively small number of concepts, form a quite large language in the end. Nevertheless, the system model proved to be adequate for our purposes. While the system model seems not free of intricacy, true is that an all-embracing language with the complexity that characterizes UML, demands a semantic domain up to it. The insights gained thanks to the effort of giving semantics to class diagrams –among others– by means of the system model, too, motivated a number of suggestions for the improvement of the system model. These will be incorporated in a new version of the system model, to be brought out in the near future.

Technische Universität Braunschweig
Informatik-Berichte ab Nr. 2003-10

| | | |
|---|---|---|
| 2003-10 | M. Mutz, M. Huhn | Automated Statechart Analysis for User-defined Design Rules |
| 2004-01 | T.-P. Fries, H. G. Matthies | A Review of Petrov-Galerkin Stabilization Approaches and an Extension to Meshfree Methods |
| 2004-02 | B. Mathiak, S. Eckstein | Automatische Lernverfahren zur Analyse von biomedizinischer Literatur |
| 2005-01 | T. Klein, B. Rumpe, B. Schätz (Herausgeber) | Tagungsband des Dagstuhl-Workshop MBEES 2005: Modellbasierte Entwicklung eingebetteter Systeme |
| 2005-02 | T.-P. Fries, H. G. Matthies | A Stabilized and Coupled Meshfree/Meshbased Method for the Incompressible Navier-Stokes Equations — Part I: Stabilization |
| 2005-03 | T.-P. Fries, H. G. Matthies | A Stabilized and Coupled Meshfree/Meshbased Method for the Incompressible Navier-Stokes Equations — Part II: Coupling |
| 2005-04 | H. Krahn, B. Rumpe | Evolution von Software-Architekturen |
| 2005-05 | O. Kayser-Herold, H. G. Matthies | Least-Squares FEM, Literature Review |
| 2005-06 | T. Mücke, U. Goltz | Single Run Coverage Criteria subsume EX-Weak Mutation Coverage |
| 2005-07 | T. Mücke, M. Huhn | Minimizing Test Execution Time During Test Generation |
| 2005-08 | B. Florentz, M. Huhn | A Metamodel for Architecture Evaluation |
| 2006-01 | T. Klein, B. Rumpe, B. Schätz (Herausgeber) | Tagungsband des Dagstuhl-Workshop MBEES 2006: Modellbasierte Entwicklung eingebetteter Systeme |
| 2006-02 | T. Mücke, B. Florentz, C. Diefer | Generating Interpreters from Elementary Syntax and Semantics Descriptions |
| 2006-03 | B. Gajanovic, B. Rumpe | Isabelle/HOL-Umsetzung strombasierter Definitionen zur Verifikation von verteilten, asynchron kommunizierenden Systemen |
| 2006-04 | H. Grönniger, H. Krahn, B. Rumpe, M. Schindler, S. Völkel | Handbuch zu MontiCore 1.0 - Ein Framework zur Erstellung und Verarbeitung domänenspezifischer Sprachen |
| 2007-01 | M. Conrad, H. Giese, B. Rumpe, B. Schätz (Hrsg.) | Tagungsband Dagstuhl-Workshop MBEES: Modellbasierte Entwicklung eingebetteter Systeme III |
| 2007-02 | J. Rang | Design of DIRK schemes for solving the Navier-Stokes-equations |
| 2007-03 | B. Bügling, M. Krosche | Coupling the CTL and MATLAB |
| 2007-04 | C. Knieke, M. Huhn | Executable Requirements Specification: An Extension for UML 2 Activity Diagrams |
| 2008-01 | T. Klein, B. Rumpe (Hrsg.) | Workshop Modellbasierte Entwicklung von eingebetteten Fahrzeugfunktionen, Tagungsband |
| 2008-02 | H. Giese, M. Huhn, U. Nickel, B. Schätz (Hrsg.) | Tagungsband des Dagstuhl-Workshop MBEES: Modellbasierte Entwicklung eingebetteter Systeme IV |
| 2008-03 | R. van Glabbeek, U. Goltz, J.-W. Schicke | Symmetric and Asymmetric Asynchronous Interaction |
| 2008-04 | M. V. Cengarle, H. Grönniger B. Rumpe | System Model Semantics of Statecharts |
| 2008-05 | M. V. Cengarle, H. Grönniger B. Rumpe | System Model Semantics of Class Diagrams |